\documentclass[aps,pre,preprint,groupedaddress]{revtex4}
\usepackage{graphics}
\usepackage{epsfig}
\begin{document}
\title{Diffusion, Fragmentation and Coagulation Processes:
Analytical and Numerical Results.}
\author{Poul Olesen, Jesper Ferkinghoff-Borg, Mogens H. Jensen, Joachim Mathiesen}

\email[]{polesen@nbi.dk,borg@nbi.dk,mhjensen@nbi.dk,mathies@nbi.dk}
\affiliation{The Niels Bohr Institute, Blegdamsvej 17, DK-2100
Copenhagen. Denmark}
\date{\today}
\pacs{82.30.Lp,68.35.Fx,68.43.Jk}
\begin{abstract}
We formulate dynamical rate equations for physical processes driven
by a combination of diffusive growth, size fragmentation and
fragment coagulation. Initially, we consider
processes where coagulation is absent. In this case we solve the
rate equation exactly leading to size distributions of Bessel type which
fall off as $\exp(-x^{3/2})$ for large $x$-values. Moreover, we
provide explicit formulas for the expansion coefficients in terms of Airy functions.
Introducing the coagulation term, the full non-linear model is mapped exactly
onto a Riccati equation that enables us to derive various asymptotic
solutions for the distribution function. In particular, we find a standard
exponential decay, $\exp(-x)$, for large $x$, and observe a crossover
from the Bessel function for intermediate values of $x$.
These findings are checked by numerical 
simulations and we find perfect agreement between the theoretical predictions and numerical results.
\end{abstract}

\pacs{}
\maketitle

\section{Introduction}
Since the pioneering work of Smoluchowski
\cite{smoluchowski1916,smoluchowski1917} dating back to the
beginning of the last century, the literature on coagulation and
fragmentation processes has grown considerably. Smoluchowski
original coagulation equation
\cite{smoluchowski1916,smoluchowski1917} provides a mean field
description of clusters that coalesce by binary collisions with a
constant rate. Scaling theory and exactly solvable models in the kinetics of irreversible
aggregation has recently been reviewed in \cite{leyvraz}.

Fragmentation and coagulation was first considered as
combined processes in \cite{melzak}, and mean field type of
coagulation-fragmentation models have subsequently been used in a
diverse range of applications, including polymer kinetics
\cite{ziff}, aerosols \cite{frielander}, cluster formation in
astrophysics \cite{silk} and animal grouping in biology
\cite{okubo,gueron1995}. We refer to \cite{drake,lauren2004} (and
references therein) for a survey of the progress in the study of
coagulation-fragmentation process. 

Recently, we have suggested a mean field model describing the
dynamics of coherent structures, like ice crystals and structural
elements of biomolecules, which grow or shrink randomly due to the
diffusive motion of their boundaries and are subject to occasional
fragmentation \cite{ferkinghoff,mathies}. In this paper, we present
an extension of the model to account for coagulation processes as
well. The extension can be considered as a generalization of the
Smoluchowski coagulation-fragmentation equation to include
processes where size diffusion is important. We emphasize that our
approach differs from the approach in e.g. \cite{inhomogene}, where the
clusters represent particles immersed in a gas or liquid type of
medium and the diffusive term added to the coagulation-fragmentation
equation represents the random movement of the center of mass of
each cluster. In contradistinction, we are focusing on systems 
where diffusion operates in the size space rather than in real
space. 

Diffusive growth processes are encountered in numerous physical 
systems. For instance, the process behind grain growth in ice or metallurgical
systems can effectively be described as a size diffusive process, 
where the diffusion constant depends on the surface tension and
the mobility of the grain boundaries (in the absence of 
extrinsic drag forces resulting from i.$\!$ e. impurities in
the material), see i.$\!$ e. \cite{mathies,ferkinghoff,hillert,louat,alley1,alley2}. 
Size diffusion also appears in natural conjunction with
convective growth in systems where the clusters in concern are coupled to a medium mediating the
addition or subtraction of monomeric units. 
Crystal growth dynamics by reversible solute addition has for instance been considered in
\cite{mccoy}, in the limit of no coagulation and
fragmentation. The combined process of fragmentation, size
diffusion and convection has been studied in \cite{flyvbjerg1,flyvbjerg2} to
describe the abrupt transitions from the growing to the shrinking state
of microtubules. In both cases, the effective convective and
diffusive growth can be directly related to the rate equations for monomer 
attachment/detachment from and to the medium (see i.$\!$ e. \cite{mccoy}).
Finally, structural transitions in polypeptide systems
\cite{zimm-bragg} can be modeled by a continuous 
master equation with a size diffusive term representing the random 
fluctuations of the ordered-disordered interface \cite{ferkinghoff}. 

Although we are not aware of previous work where diffusive growth appears
in conjunction with cluster coagulation, we believe --in light of the 
examples given above-- that such extension to Smoluchowskis original 
fragmentation-coagulation model is natural and should find
application in several physical systems. Our aim in this paper is
therefore to elucidate the effect of size diffusion in simple
models of coagulation-fragmentation processes.

To proceed, let $N(x,t)$ denote the number of clusters of size $x$ at time $t$.
The general form of the coagulation-fragmentation model then reads \cite{drake}
\begin{equation}
\partial_t N(x,t) = [\partial_t N]_{\mbox{coag}}+[\partial_t
N]_{\mbox{frag}}.
\end{equation}
where
$$
[\partial_t N]_{\mbox{coag}}=\frac{1}{2}\int_0^x K(x-x',x')N(t,x-x')N(t,x')dx'-
N(t,x)\int_0^\infty K(x,x')N(t,x')dx'
$$
and
$$
[\partial_t
N]_{\mbox{frag}}=-N(x,t)\int_0^x F(x-x',x')dx' +2\int_0^\infty F(x,x')N(x+x',t)dx'.
$$
The microscopic details of a given physical model is embedded in the
kernels $K(x,x')\ge 0$ and $F(x,x')\ge 0$ (symmetric in $x$ and $x'$) which
represent, respectively, the rate of coagulating two clusters of size $x$ and $x'$ into one cluster of size $x+x'$ and the rate of fragmentation of a
cluster of size $x+x'$ into two clusters of size $x$ and $x'$.
The possible exchange of monomeric units with an external medium
can be modeled by adding a convective and diffusive term to 
the equation which then reads,
\begin{equation}
\label{eq0}
\partial_t N(x,t) = -v(x)\partial_x N(x,t)+D(x)\partial^2_x N(x,t)+[\partial_t
N]_{\mbox{coag}}+[\partial_t N]_{\mbox{frag}}.
\end{equation}
Here, the first term describes the average drift of the cluster
sizes due to monomeric exchange with the medium and the second term represents the random fluctuations
superimposed on this drift (size diffusion). In the following we will examine analytical and numerical aspects of
this equation with the simple choice of size-­independent parameters;
$D(x)=D$, $F(x,x')=f$ and $K(x,x')=\beta$ and setting $v=0$. The model
presented in \cite{ferkinghoff,mathies} for the dynamics of ice crystals and
structural building blocks in biomolecules, corresponds to case of no
coagulation $\beta=0$. In these cases the systems are considered
closed, so the $v=0$ condition follows from excluding any external driving. When the system
is in a solution allowing for the solute-solvent exchange of monomeric units
(as in i.$\!$ e. \cite{mccoy,flyvbjerg1,flyvbjerg2}) the $v=0$
condition corresponds to a situation where the solute (clusters) are in
thermodynamical equilibrium with the solvent. The existence of steady
states in the model with no coagulation has recently been identified for a wider class of fragmentation
kernels \cite{laurencot2004}, where connections to the pure
fragmentation equation \cite{ernst} and the so-called
``shattering'' transition \cite{shatter} related to the formation
of dust particles, is further discussed. In the absence of diffusion, the model
has been applied to the kinetics of reacting polymers
in \cite{bak,stewart}, where the uniqueness of solutions and convergence
to equilibrium in the limit $t\rightarrow  \infty$ have been proved.
As will be demonstrated, the inclusion of a diffusion term in the
equation may lead to solutions radically different from the ones found in
\cite{bak,stewart}. 

Our paper is organized as follows. In section 2, we derive the exact
solution to the model in the case of no coagulation
($D>0,~f>0,~\beta=0$). Section 3 and 4 is devoted to the discussion
of the expansion coefficients entering this solution. In particular,
we give an explicit formula for the coefficients in terms of a
orthogonal set of Airy functions and discuss their asymptotic
behavior. A moment analysis of the solution in the large time limit
is carried out in section 5. In section 6, we focus on the full
diffusion-coagulation-fragmentation model. It is shown that the
equation can be mapped to a Riccati equation, and we discuss its
solutions in the case of no coagulation ($\beta=0$, section 7) and
no diffusion ($D=0$, section 8). Here, the solution to the
diffusion-fragmentation model found in section 2-5 becomes useful to
elucidate the structure of the solutions to the Riccati equation. In
the section 9 we return to the full
diffusion-fragmentation-coagulation model. We demonstrate that the
equation possesses a transition point for the coagulation term,
where the distribution crosses over from the Bessel behavior found in the
pure fragmentation-diffusion case to an exponential. In the final section,
we discuss two numerical implementations of the model.

\section{Solution to the fragmentation-diffusion equation}
In the following three sections, we expand on the results of a
recent letter \cite{ferkinghoff} on the diffusion-fragmentation
equation. In particular, we shall carefully study the structure of
the analytical solution of the diffusion-fragmentation equation. We
shall derive orthogonality relations of the functions entering the
solution and find the corresponding expansion coefficients.
The general equation we wish to solve has the form
$$
\partial_t N(x,t)=D\partial^2_xN(x,t)-fxN(x,t)+2f\int_x^\infty N(x',t)dx',
$$
which is a particular case of Eq. (\ref{eq0}) with size independent
diffusion and fragmentation terms, and with no coagulation
($\beta=0$). Below we shall consider the boundary conditions $N(0,t)=0$
and $N(x,t)\rightarrow 0$ for $x\rightarrow\infty$ for all
times $t$, including the initial condition at $t=0$.

\begin{figure}
\epsfig{width=.6\textwidth,file=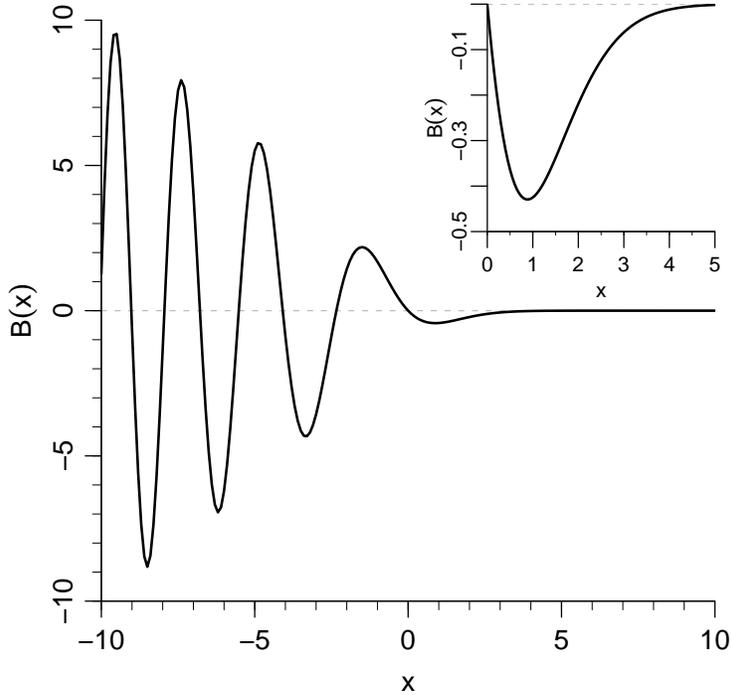}
\caption{The function $B(x)$ ($x$ is dimensionless) and its intersections with
  $y=0$. Note that all intersections appears for $x<0$ only. Inset: A zoom of the function for positive $x$.}\label{fig1}
\end{figure}

We rescale time and the cluster size
$$
x\rightarrow x/x_0~~{\rm and}~~t\rightarrow t/(fx_0)^{-1},
$$
where $x_0=(D/f)^{1/3}$. A subsequent differentiation with respect
to $x$ leads to the equation:
\begin{equation}
\label{first}
\partial_t\partial_x N(x,t)= \partial^3_x N(x,t)-3N(x,t)-x\partial_x N(x,t).
\end{equation}
It is fairly easy to solve Eq. (\ref{first}) by separation of
variables. Taking $N(x,t)=T(t)X(x)$ we have
\begin{equation}
\dot{T}(t)=\lambda T(t),~~X'''(x)-3X(x)-xX'(x)=\lambda X'(x),
\label{2}
\end{equation}
where $\lambda$ is a separation constant. We shall simplify the
notation by shifting the variable $x$ such that the separation
constant gets included, $x\rightarrow x-\lambda$. The equation for
$X$ can be solved (with the boundary condition $X(x)\rightarrow 0$ for
$x\rightarrow\infty$) by means of a Fourier transform, with the
result 
\begin{equation}
X(x)=\frac{1}{2}\int_{-\infty}^\infty ~dk~ k^2~e^{ikx+ik^3/3}\equiv B(x).
\label{B}
\end{equation}
Although $x$ is positive we need to consider $X(x)$ for negative
values of $x$ in equation (\ref{solution_first}). By analytical
continuation the function $X(x)$ is well defined for $x<0$. The
function $B$ is related to the Airy integral  
\begin{equation}
A(x)\equiv \frac{1}{2}\int_{-\infty}^\infty ~dk~e^{ikx+ik^3/3}=\int_0^\infty~
dk~\cos \left(kx+\frac{k^3}{3}\right),
\label{A}
\end{equation}
which has been discussed e.$\!$g. in reference \cite{watson}, see in
particular pp.188-190. In the panels of Fig. \ref{fig1} we show $B(x)$
for positive and negative values of $x$, respectively.

By differentiating $A(x)$ twice we get
\begin{equation}
B(x)=-\partial_x^2~A(x).
\label{B-A}
\end{equation}
Collecting the results obtained above, and inserting once again the
separation constant, the solution of Eq. (\ref{first}) can now be written
\begin{equation}
N(x,t)=\sum_n ~C_ne^{\lambda_n t}B(x+\lambda_n).
\label{solution_first}
\end{equation}
In case the eigenvalues $\lambda_n$ are in a continuous range the sum
should be replaced by an integral. We shall here impose an absorbing
boundary conditions for small clusters, which implies that the
probability of clusters with zero size vanishes at all times, i.$\!$ e.
\begin{equation}
N(0,t)=0.
\label{boundary}
\end{equation}
We implement this condition by requiring $B(\lambda_n)=0$. Thus the
eigenvalues must be zeros of the function $B(x)$.

To find possible zeros of $B(x)$ we need some properties of the
functions $A(x)$ and $B(x)$. It turns out that $A(x)$ can be
expressed in terms of Bessel functions \cite{watson} by use of a
deformation of the contour in Eq. (\ref{A}), with the result
\begin{eqnarray}
A(x)&=&\sqrt{\frac{x}{3}}~K_{\frac{1}{3}}\left(\frac{2x^{3/2}}{3}\right)~~{\rm
for}~~x>0\nonumber \\
&=&\frac{\pi}{3}\sqrt{|x|}\left[J_{\frac{1}{3}}\left(\frac{2|x|^{3/2}}{3}
\right)+
J_{-\frac{1}{3}}\left(\frac{2|x|^{3/2}}{3}\right)\right]~~{\rm for}~~x<0.
\label{bessel}
\end{eqnarray}
The absolute signs in the last line should be noticed. The function $A(x)$ is
positive for $x>0$ and oscillates for $x<0$, where it has an infinite
number of zeros.

We can now differentiate $A(x)$, and using well
known functional relations\footnote{We use $zJ'_\nu (z)+\nu J_\nu (z)=z
J_{\nu -1}(z)$ and $ zJ'_\nu (z)-\nu J_\nu (z)=-zJ_{\nu +1}(z)$, and the
relations for the $K-$functions, $K'_\nu (z)=-(\nu/z)K_\nu (z)-K_{\nu -1} (z)$
and $K_\nu (z)=K_{-\nu}(z).$ } for the
Bessel functions \cite{watson,nielsen}, one obtains the result
\begin{eqnarray}
A'(x)&=&-\frac{x}{\sqrt{3}}K_{\frac{2}{3}}\left(\frac{2x^{3/2}}{3}\right)~~{\rm
for}~~x>0,\nonumber \\
&=&\frac{\pi}{3} x\left[J_{-\frac{2}{3}}\left(\frac{2|x|^{3/2}}{3}\right)
-J_{\frac{2}{3}}\left(\frac{2|x|^{3/2}}{3}\right)\right]~~{\rm for}~~x<0.
\label{A'}
\end{eqnarray}
Performing another differentiation gives
\begin{eqnarray}
A''(x)&=&x\sqrt{\frac{x}{3}}~K_{\frac{1}{3}}\left(\frac{2x^{3/2}}{3}\right)~~
{\rm for}~~ x>0, \nonumber \\
&=&\frac{\pi}{3}x\sqrt{-x}\left[J_{\frac{1}{3}}\left(\frac{2|x|^{3/2}}{3}
\right)+
J_{-\frac{1}{3}}\left(\frac{2|x|^{3/2}}{3}\right)\right]~{\rm for}~~x<0.
\end{eqnarray}
Comparison with (\ref{bessel}) shows that $A(x)$ satisfies the simple second
order differential equation found by Stokes \cite{watson}
\begin{equation}
\partial^2_x A(x)=xA(x),
\label{differential}
\end{equation}
which will play a crucial role in the following. Comparing this equation to
(\ref{B-A}) we see the remarkably simple relation between $B(x)$ and $A(x)$,
\begin{equation}
B(x)=-xA(x).
\label{crucial}
\end{equation}

From well known properties of the Bessel functions $K$ and $J$ we see that
the function $A(x)$ has zeros for $x=\lambda_n<0$ with
\begin{equation}
J_{\frac{1}{3}}\left(\frac{2|\lambda_n|^{3/2}}{3}
\right)+
J_{-\frac{1}{3}}\left(\frac{2|\lambda_n|^{3/2}}{3}\right)=0.
\label{zeros}
\end{equation}
These zeros can be computed easily numerically. The first few are given
approximately by
\begin{equation}
\lambda_1=-2.338,~\lambda_2=-4.088,~\lambda_3=-5.521,~\lambda_4=-6.787,~\lambda
_5=-7.945,~\lambda_6=-9.023.
\end{equation}
\vspace{0.3cm}
It should be mentioned that from the asymptotic form of the Bessel functions
the zeros are approximately given by
\begin{equation}
\lambda_n\approx -\left[\frac{3\pi}{2}\left(n+\frac{3}{4}\right)\right]^{2/3}.
\label{asymptotic}
\end{equation}
This expression is valid for large $n$. However, the expression
provides a surprisingly accurate estimate of $\lambda_{n+1}$ even
for small $n'$s. For example, for $n=0$ one finds that
Eq.(\ref{asymptotic}) gives $-2.32$, as compared to the numerically
obtained more accurate value $\lambda_1=-2.338$. For $n=3$ one
obtains $\approx -6.785$ to be compared to the more accurate
$\lambda_4=-6.787$.

We now return to the boundary condition (\ref{boundary}) which is to be
implemented on the solution (\ref{solution_first}). This requires
\begin{equation}
N(0,t)=\sum_{n=0}^\infty ~C_n B(\lambda_n)=0.
\label{boundary2}
\end{equation}
Taking into account Eq. (\ref{crucial}) giving the relation between
$A(x)$ and $B(x)$ as well as the expression (\ref{bessel}) for
$A(x)$, we thus see that the $\lambda_n'$s are discrete (since there
are only discrete zeros in the Bessel functions), they must be
negative and must also satisfy Eq. (\ref{zeros}). However, even
though $A(0)\neq 0$ it follows from Eq. (\ref{crucial}) that
$B(0)=0$. Therefore the sum in Eq. (\ref{boundary2}) includes $n=0$
with $\lambda_0=0$.

The solution to (\ref{first}) is thus
\begin{equation}
N(x,t)=C_0~B(x)+\sum_{n=1}^\infty ~C_n e^{\lambda_nt}B(x+\lambda_n).
\label{fullsolution}
\end{equation}
Here we separated the first term explicitly in order to emphasize that
it is time independent. Since the $\lambda_n'$s are negative for $n\neq 0$
it follows that after a sufficiently long time the first term dominates,
\begin{equation}
N(x,t)\rightarrow C_0~B(x)~~{\rm for}~~t\rightarrow\infty.
\end{equation}
Please note that $C_0<0$. In order to use Eq. (\ref{fullsolution}) we need to determine the
constants $C_n$. This problem will be discussed in the next section.

\section{Orthogonality}

The solution (\ref{fullsolution}) is incomplete as it stands since we need to
determine the expansion coefficients. This problem is related to orthogonality
of the functions entering this solution. It turns out that the
$B(x+\lambda_n)'$s are not orthogonal\footnote{This can easily be seen
in examples by doing the relevant integrals numerically. Thus, for example,
the integral $\int_0^\infty dx B(x-2.338)B(x-4.088)$ has the non-vanishing
value $-9.102$.}, which is connected to the fact
that the $B-$function satisfies the third order differential equation
(\ref{first}). However, we shall now use the fact that $A(x)$ satisfies the
second order differential equation (\ref{differential}), so we can use
standard orthogonality consideration as far as $A(x)$ is concerned. Therefore
instead of $N(x,t)$ we consider the function
\begin{equation}
M(x,t)=\sum_{n=0}^\infty C_n e^{\lambda_n t}A(x+\lambda_n),
\label{expansion}
\end{equation}
where the constants $C_n$ are the same as in Eq.
(\ref{solution_first}). The connection between $N(x,t)$ and $M(x,t)$
is given by
\begin{equation}
\partial_x^2~M(x,t)=-N(x,t)
\label{v-u}
\end{equation}
as a consequence of Eq. (\ref{B-A}).

We can now show that the functions $A(x+\lambda_n)$ form an
orthogonal set of functions for $\lambda_n\neq 0$. From Eq.
(\ref{differential}) we have
\begin{equation}
A''(x+\lambda_n)=(x+\lambda_n)A(x+\lambda_n).
\end{equation}
This second order equation allows us to use standard methods, in contrast
to the original third order equation for $N(x,t)$. Thus we obtain
\begin{equation}
\int_0^\infty dx\left[A(x+\lambda_m)A''(x+\lambda_n)-A(x+\lambda_n)
A''(x+\lambda_m)\right]=(\lambda_n-\lambda_m)\int_0^\infty~dx A(x+\lambda_n)
A(x+\lambda_m).
\label{ortho}
\end{equation}
Following standard procedures we now shift the two differentiations in the
first term in the integrand on the left hand side by two partial
differentiations. In this way the first term cancels the second, except for
contributions from the boundaries. For $x=\infty$ there are no contributions,
since the Bessel function vanishes exponentially. Taking into account the
contributions from the lower limit $x=0$ we have
\begin{equation}
(\lambda_n-\lambda_m)\int_0^\infty~dx A(x+\lambda_n)A(x+\lambda_m)=
-A(\lambda_m)A'(\lambda_n)+A'(\lambda_m)A(\lambda_n).
\label{lambda17}
\end{equation}
However, since $\lambda_n$ are the zeros of the function $A$ for
$n\neq 0$, we have $A(\lambda_m)=A(\lambda_n)=0$. It is important to
notice that neither $A(0)$ nor $A'(0)$ vanish. For $n\neq 0$ the
quantity on the right hand side of Eq. (\ref{lambda17}) vanishes,
and hence
\begin{equation}
(\lambda_n-\lambda_m)\int_0^\infty~dx A(x+\lambda_n)A(x+\lambda_m)=0~~
{\rm for}~n~{\rm and}~m\neq 0.
\end{equation}
Thus the set $\{A(x+\lambda_n)\}$ consists of orthogonal functions,
except for $\lambda_0=0$. It then follows that the expansion
coefficients in Eq. (\ref{expansion}) are determined through the
equation\footnote{If the initial data are taken to be given at time
$t_0$ instead of $t=0$, $C_m$ should be replaced by
$C_me^{\lambda_nt_0}$ and $M(x,0)$ by $M(x,t_0)$ in Eq.(\ref{C}).
Similarly, in the solution $e^{\lambda_nt}$ is replaced by
$e^{\lambda_n(t-t_0)}$} \cite{ferkinghoff,mathies}
\begin{equation}
C_m=\frac{1}{I_m}~\left[\int_0^\infty~dx~M(x,0)A(x+\lambda_m)-C_0\int_0^\infty
dxA(x+\lambda_m)A(x)\right],~~m\neq 0,
\label{C}
\end{equation}
where
\begin{equation}
I_m=\int_0^\infty~dx~A(x+\lambda_m)^2.
\label{I}
\end{equation}
In the next section we shall show the $I_m$ can be expressed in terms of
$A'(\lambda_n)$, which in turn can be expressed in terms of Bessel
functions \cite{watson}.

Taking $\lambda_m=0$ we obtain from Eq. (\ref{lambda17})
\begin{equation}
\int_0^\infty dxA(x+\lambda_n)A(x)=-\frac{A(0)A'(\lambda_n)}{\lambda_n},
\label{747}
\end{equation}
which we shall use in the next section.

In the expression (\ref{C}) for $C_n$ the function $M(x,0)$ enters.
However, the relevant initial function is $N(x,0)$. We can reexpress
$C_m$ in terms of $N(x,0)$ by solving Eq. (\ref{v-u}) for $M$ in
terms of $N$. Noticing that the Greens function in one dimension is
$\theta (x)x$, we easily obtain from (\ref{v-u})
\begin{equation}
M(x,0)=\alpha +\beta x-\int_0^x~dx'~(x-x')N(x',0),
\label{18}
\end{equation}
where $\alpha$ and $\beta$ are integration constants. It follows from
the definition, Eq. (\ref{expansion}), of $M$ that $M(0,0)=C_0A(0)$, since
$A(\lambda_n)=0$ for $n\neq 0$. Thus $\alpha=C_0A(0)$.

To find the constant $\beta$ consider
\begin{equation}
\partial_x M(x,0)|_{x=0}=\beta=\sum_{n=0}^\infty ~C_nA'(\lambda_n).
\end{equation}
Now from the solution (\ref{solution_first}) we have by use of (\ref{B-A})
\begin{equation}
\int_0^\infty ~ dx~N(x,0)=\sum_{n=0}^\infty C_n\int_0^\infty dx B(x+\lambda_n)
=-\sum_{n=0}^\infty C_n\int_0^\infty dx \partial_x^2A(x+\lambda_n)=
\sum_0^\infty C_nA'(\lambda_n).
\label{29}
\end{equation}
Therefore
\begin{equation}
\beta=\int_0^\infty ~ dx~N(x,0).
\end{equation}
We now insert these results in Eq.(\ref{18}) which becomes
\begin{equation}
M(x,0)=-\int_0^x~dx~(x-x')~N(x',0)+x\int_0^\infty~dx'~N(x',0)+C_0A(0).
\label{20}
\end{equation}
To proceed we need also to determine $C_0$. We have by use of Eqs.
(\ref{solution_first}) and (\ref{B-A})
\begin{equation}
\int_0^\infty ~ dx~xN(x,0)=\sum_{n=0}^\infty C_n\int_0^\infty dx~x
B(x+\lambda_n)
=-\sum_{n=0}^\infty C_n\int_0^\infty dx x\partial_x^2A(x+\lambda_n)=
-C_0~A(0),
\label{32}
\end{equation}
so
\begin{equation}
C_0=-\frac{1}{A(0)}~\int_0^\infty ~ dx~xN(x,0),~~A(0)=\frac{\pi}{3^{2/3}\Gamma
(2/3)}.
\label{c0}
\end{equation}
Thus $C_0$ is determined in terms of the initial data for $N$.

\begin{figure}
\epsfig{width=.5\textwidth,file=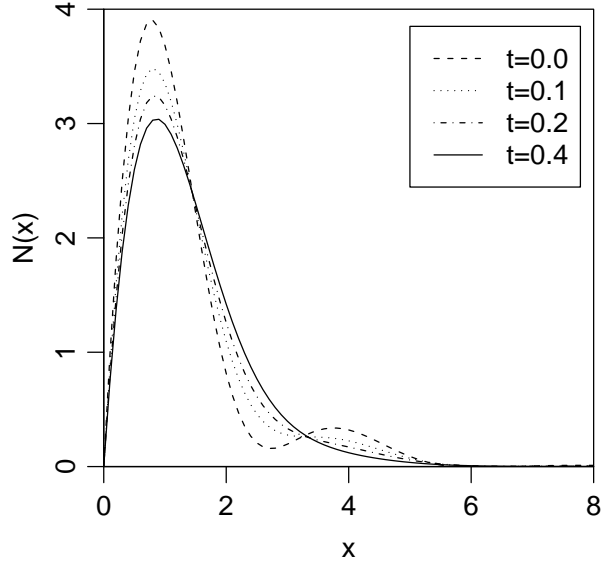}
  \caption{An example showing
the time development of $N(x,t)$ ($x$ and $t$ are dimensionless)  with a secondary peak, which
ultimately disappears. The specific values of the expansion
parameters $C_n$ are $(C_0,C_1,C_2 \ldots)=(-7,-.2,-.2,.05,-.035)$
.}\label{expcoef}
\end{figure}

We can now insert the result (\ref{c0}) in Eq. (\ref{20}),
\begin{equation}
M(x,0)=\int_x^\infty~dx'~(x-x')N(x',0).
\label{34}
\end{equation}
This finally gives an expression for the expansion coefficients in terms
of the initial value for $N$,
\begin{equation}
C_m=\frac{1}{I_m}~\int_0^\infty dx~A(x+\lambda_m)\left[\int_x^\infty~dx'~
(x-x')N(x',0)-C_0~A(x)\right],
\label{cm}
\end{equation}
where we inserted the solution (\ref{34}) of $M$ in terms of $N$ in
the expression (\ref{C}) for $C_m$. In Fig. \ref{expcoef} we show an
example of the evolution of $N(x,t)$ with the interesting feature of
the presence of a secondary peak.
Interchanging the $x$ and $x'$ integrations this expression can be
rewritten in a form which is more suitable for insertions of data
for $N(x,0)$, namely
\begin{eqnarray}
C_m&=&\frac{1}{I_m}~\left[\int_0^\infty ~dx~N(x,0)\int_0^x ~dx'~(x'-x)
A(x'+\lambda_m)-C_0~\int_0^\infty ~dx~A(x+\lambda_m)A(x)\right]\nonumber \\
&=&\frac{1}{I_m}~\int_0^\infty ~dx~N(x,0)\left\{A'(x+\lambda_m)-
A'(\lambda_m)-(x+\lambda_m)\int_0^xdx'~A(x'+\lambda_m)\right.
\nonumber \\
&&\left.+\frac{x}{A(0)}\int_0^\infty dx' ~A(x'+\lambda_m)A(x')\right\}.
\label{cmm}
\end{eqnarray}
To obtain the last form we used Eq. (\ref{B-A}) and inserted $C_0$.
Thus, to sum up the solution of Eq. (\ref{1}) is given by Eq.
(\ref{solution}), with $B$ given by Eqs. (\ref{crucial}) and
(\ref{bessel}), and with $C_n$ given by Eq. (\ref{cmm}). It should
be emphasized that in Eq. (\ref{cmm}) the initial function $N(x,0)$
enters only in the first integral, whereas the other integrals
involving the function $A$ can be determined numerically to any
desired accuracy.

\section{On the expansion coefficients}

In this section we shall examine the expansion coefficients $C_m$
given by Eqs. (\ref{cm}) and (\ref{cmm}). We start by considering
the normalization integrals $I_m$ defined by (\ref{I}). The
differential equation (\ref{differential}) for $A(x)$, $A''=xA$, can
be rewritten as $2A'A''-2xAA'-A^2=-A^2$, i.$\!$ e.
\begin{equation}
\frac{d}{dx}~\left(A'(x)^2-xA(x)^2\right)=-A(x)^2,
\end{equation}
so that the normalization integral becomes
\begin{equation}
I_m=\int_{\lambda_m}^\infty dx A(x)^2=A'(\lambda_m)^2,
\end{equation}
where we used that $A(\lambda_m)=0$.

We can now express $I_m$ in terms of Bessel functions by
differentiating $A(x)$ given in Eq. (\ref{bessel}) and using
standard functional relations for the relevant Bessel functions
\cite{watson,nielsen}. Using $A'(x)$ already evaluated in Eq.
(\ref{A'}) we have
\begin{equation}
I_m=\frac{\pi^2}{9}~\lambda_m^2~[J_{-\frac{2}{3}}(2|\lambda_m|^{3/2}/3)-
J_{\frac{2}{3}}(2|\lambda_m|^{3/2}/3)]^2.
\label{pp}
\end{equation}
If we use the well known asymptotic expansions of the Bessel
functions as well as the asymptotic eigenvalue (\ref{asymptotic}) it
is easy to see that Eq. (\ref{pp}) becomes
\begin{equation}
I_m\approx \pi^{4/3}\left(\frac{3}{2}\right)^{1/3}~\left(m+\frac{3}{4}
\right)^{1/3},~~
m\rightarrow\infty.
\label{as}
\end{equation}
We have compared (\ref{as}) with the exact expression (\ref{pp}) even
for small $m$. We find that for $m=0$ the analytic expression (\ref{pp}) gives
4.85285, whereas the approximate result (\ref{as}) gives 4.785. For $m=4$ the
corresponding numbers are 8.85737 and 8.8538. For $m\geq 5$ one gets the
first two decimals right.

The first expression (\ref{cmm}) can now be simplified by use of Eq.
(\ref{747}),
\begin{equation}
C_m=\frac{1}{A'(\lambda_m)^2}\int_0^\infty dx N(x,0)\int_0^x dx' (x'-x)
A(x'+\lambda_m)-\frac{1}{\lambda_mA'(\lambda_m)}\int_0^\infty dx xN(x,0),
~~m\neq 0.
\label{757}
\end{equation}
From the asymptotic statements (\ref{asymptotic}) and (\ref{as}) we
see that the last term on the right hand side of (\ref{757}) behaves
like $m^{-5/6}$. However, in the expansion for $M$ this is to be
multiplied by the oscillating function $A(x+\lambda_n)$. In any case
it should be noticed that if $t\neq 0$ there is no convergence
problem because  the exponential damping factors $e^{\lambda_m t}$
give rapid convergence. Convergence problems in the form of very
slow convergence may arise at the initial time, where the expansion
coefficients are determined in terms of the initial data.

\section{Mean values}

We shall now compute the mean values $\overline{x}$ and $\overline{x^2}$,
which turn out to be given by rather simple expressions for large times. We
have
\begin{equation}
\overline{x}=\int_0^\infty dx ~xN(x,t)\left/\int_0^\infty dx~N(x,t)
\right.,~~{\rm and}~~
\overline{x^2}=\int_0^\infty dx ~x^2N(x,t)\left/\int_0^\infty dx~N(x,t)\right..
\end{equation}
By the methods already used in Eqs. (\ref{29}) and (\ref{32}) we
obtain
\begin{equation}
\overline{x}=-A(0)\left/\left(A'(0)+(1/C_0)\sum_{n=1}^\infty C_ne^{\lambda_nt}
A'(\lambda_n)\right)\right..
\end{equation}
We also have by use of Eq. (\ref{B-A}) and two partial integrations
\begin{equation}
\int_0^\infty dx~x^2 N(x,t)=-2\sum_{n=0}^\infty C_ne^{\lambda_nt}
\int_0^\infty dx~A(x+\lambda_n).
\end{equation}
In the limit $t\rightarrow\infty$ we obtain the simple results
\begin{equation}
\overline{x}\rightarrow \frac{3^{2/3}~\Gamma(4/3)}{\Gamma (2/3)}\approx 1.3717,
\end{equation}
as well as
\begin{equation}
\overline{x^2}\approx 2.5758.
\end{equation}
Here we used that $\int_0^\infty dx~A(x)\approx 1.0472$.
The relative dispersion therefore becomes
\begin{equation}
\frac{D^2}{\overline{x}^2}\equiv \frac{\overline{x^2-\overline{x}^2}}
{\overline{x}^2}\approx 0.36897.
\end{equation}
for $t\rightarrow\infty$. The dispersion is therefore
smaller than $\overline{x}$, $D\approx 0.60743 ~\overline{x}$.

\section{Transformation of the model with finite coagulation to a Riccati
equation}

In this section we shall discuss the model, Eq. (\ref{eq0}), with
constant diffusion and finite size-independent fragmentation and
coagulation kernels, $f>0$ and $\beta>0$. The basic equation for the steady state solution is thus \cite{lushnikov}
\begin{equation}
\label{fullequation}
D~\partial_x^2 N(x)-fxN(x)+2f\int_x^\infty dx'N(x')+\beta/2\int_0^x dx'
N(x')N(x-x')-2\tilde{\beta}N(x)=0,
\end{equation}
where
\begin{equation}\label{tildebeta}
\tilde{\beta}=\beta/2 \int_0^\infty dx N(x).
\end{equation}
By integrating Eq. (\ref{fullequation}) we
easily get
\begin{equation}
\frac{D\beta}{f^2}N'(0)=\frac{\beta}{f}\int_0^\infty dx~ xN(x)-2b^2\geq 0,
\label{2b}
\end{equation}
where $b=\tilde{\beta}/f$ represents a new length scale compared to
the one given by the diffusion to fragmentation ratio,
$x_0=(D/f)^{1/3}$, as previously introduced.

Eq. (\ref{fullequation}) is an integro-differential equation. In general we
would like to solve this equation with the boundary conditions that
$N(0)=0$ and $N(x)\rightarrow 0$ for $x\rightarrow\infty$.
Furthermore, we need of course also to invoke the condition that
$N(x)$ is positive. This is not automatically guaranteed from Eq.
(\ref{fullequation}). We shall see that
if diffusion is absent, it is not possible to achieve $N(0)=0$. The
main result of this section is that in Fourier space the integro-differential
equation (\ref{fullequation}) can be mapped to a Riccati equation, or alternatively,
a linear second order differential equation.
If we can achieve $N(0)=0$ we need $N'(0)\geq 0$ from positivity of
$N(x)$, and hence the right hand side of Eq. (\ref{2b}) must be
positive. Since Eq. (\ref{2b}) is a consequence of (\ref{fullequation}) a
solution for the function $N(x)$ should satisfy (\ref{2b})
automatically.

In the presence of diffusion, we can make Eq. (\ref{fullequation})
dimensionless by rescaling the cluster size $x\rightarrow x/x_0$ as
in section II:
\begin{equation}
\partial_x^2 N(x)-xN(x)+2\int_x^\infty dx'N(x')+\frac{\beta}{2fx_0}\int_0^x dx'
N(x')N(x-x')-\frac{2b}{x_0}N(x)=0,
\label{1}
\end{equation}

We now take the Fourier transform
\begin{equation}
N(x)=\int_{-\infty+i\epsilon }^{\infty+i\epsilon } d\omega~ e^{i\omega x}
\tilde{N}(\omega )
=e^{-\epsilon x}\int_{-\infty}^{\infty} d\omega~ e^{i\omega x}
\tilde{N}(\omega ),
\label{fourier}
\end{equation}
where the convergence factor is needed in the following. Alternatively we can
consider the contour to be shifted slightly to the upper half plane. We assume
that the function $\tilde{N}(\omega )$ is continuous as the real axis is
approached.

We can now translate Eq. (\ref{1}) to Fourier space. For example, we
use
\begin{equation}
\int_x^\infty dx'~N(x')=\int_{-\infty}^\infty d\omega ~\tilde{N}(\omega )
\int_x^\infty e^{i\omega x'-\epsilon x}=i\int_{-\infty +i\epsilon}^{\infty +
i\epsilon} d\omega ~\frac{\tilde{N}(\omega )}{\omega }~e^{i\omega x}.
\end{equation}
Here the convergence factor is of course important. Putting
everything together we obtain from Eq. (\ref{1})
\begin{equation}
\int_{-\infty +i\epsilon}^{\infty +i\epsilon}d\omega ~e^{i\omega x}{\cal O}
(\omega)\tilde{N}(\omega)=\frac{\beta }{2fx_0}\int\int_{-\infty +i\epsilon}^
{\infty +i\epsilon}d\omega d\omega' ~\frac{\tilde{N}(\omega )\tilde{N}
(\omega' )}{\omega-\omega'}~\left(e^{i\omega x}-e^{i\omega' x}\right),
\end{equation}
where the operator $\cal O$ is given by
\begin{equation}
\label{operator}
{\cal O}(\omega )=-\omega^2+\frac{2i}{\omega }-\frac{2b}{x_0}-i\frac{d}{d\omega}.
\end{equation}
From this we obtain by an integration over $x$ from 0 to $\infty$
with the weight factor exp$(-i\tau x)$
\begin{equation}
i\int_{-\infty+i\epsilon}^{\infty +i\epsilon}d\omega~\frac{{\cal O}(\omega)
\tilde{N}(\omega)}{\omega-\tau}=\frac{\beta}{2fx_0}~ F(\tau)^2,
\label{3}
\end{equation}
where $F$ is the Hilbert transform
\begin{equation}
F(\tau )=\int_{-\infty+i\epsilon}^{\infty +i\epsilon}d\omega~\frac{\tilde{N}
(\omega)}{\tau-\omega}.
\label{hilbert}
\end{equation}
Eq. (\ref{3}) can be reformulated as an equation for $F$. To do this, we
rewrite $\omega^2=(\omega^2-\tau^2)+\tau^2$ to obtain\footnote{In order
to simplify the notation we leave out the limits on the $\omega-$ integration,
always assuming that this integration goes from $-\infty+i\epsilon$ to
$\infty+i\epsilon$ in the following.}
\begin{equation}
\int d\omega~\frac{\omega^2\tilde{N}(\omega)}{\omega-\tau }=\int d\omega
(\omega+\tau )\tilde{N}(\omega)+\tau^2 F(\tau)=-iN'(0)+\tau N(0)+\tau^2
F(\tau ).
\label{diffusion}
\end{equation}
We want to solve the problem with the boundary condition $N(0)=0$, so in
the following the linear term on the right hand side will be left out.
We also have
\begin{equation}
\int d\omega \frac{\tilde{N}(\omega )}{\omega (\omega-\tau)}=\frac{i\tilde
\beta }{\beta\tau}+\frac{1}{\tau }~F(\tau ).
\end{equation}
Collecting these results Eq. (\ref{3}) becomes
\begin{equation}
\frac{\beta}{2fx_0}F(\tau)^2=-\frac{dF(\tau )}{d\tau}+\left(\frac{2}{\tau}+i\tau^2+
\frac{2ib}{x_0}\right)F(\tau )-N'(0)-2i\frac{\tilde\beta}{\beta\tau}.
\label{ricatti}
\end{equation}
This is a standard equation of the Riccati type. It should be noticed that if
there is no diffusion the Riccati equation simplifies to
\begin{equation}
\frac{\beta}{2f}F(\tau)^2=-\frac{dF(\tau )}{d\tau}+\left(\frac{2}{\tau}+
2ib\right)F(\tau )-2i\frac{\tilde\beta}{\beta\tau}~~~({\rm No~~ diffusion}),
\label{nodiffusion}
\end{equation}
where $\tau$ now has the dimension of $\mbox{length}^{-1}$.

It is well known that a Riccati equation can be transformed to a linear
second order differential equation by the substitution
\begin{equation}
\frac{\beta}{2fx_0}~F(\tau )=\frac{d}{d\tau }~\log~ u(\tau ),
\label{log}
\end{equation}
where $u$ satisfies
\begin{equation}
u''(\tau )-\left(\frac{2}{\tau}+i\tau^2+
2i\frac{b}{x_0}\right)u'(\tau )+\left(\frac{\beta}{2fx_0}N'(0)+\frac{2ib}{x_0\tau}\right)u(\tau )
=0.
\label{linear}
\end{equation}

The original Fourier transform (\ref{fourier}) was given in terms of the
function $\tilde{N}$, which in turn is related to the function $F$ by the
Hilbert transform (\ref{hilbert}). However, it is well known that the
Fourier transform of a Hilbert transform has a remarkably simple property,
which in our case (since $x>0$) leads to
\begin{equation}
N(x)=\frac{1}{2\pi i}\int_{-\infty}^\infty d\tau e^{i\tau x} F(\tau ).
\label{fourierf}
\end{equation}
This equation can be derived by noticing that since the $\omega-$integration
is shifted to slightly above the real axis, we have
\begin{equation}
F(\tau)=F_P(\tau)+i\pi \tilde{N}(\tau ),~~F_P(\tau )=P\int_{-\infty}^\infty
d\omega ~\frac{\tilde{N}(\omega )}{\tau-\omega},
\label{17}
\end{equation}
where $P$ means the principal value. Further we have
\begin{equation}
N(x)=\int d\omega~\tilde{N}(\omega)~e^{i\omega x}=\frac{1}{i\pi}\int d\omega~
\tilde{N}(\omega)~e^{i\omega x}~P\int_{-\infty}^\infty dy~\frac{e^{iyx}}{y}=
\frac{1}{i\pi}\int d\tau e^{i\tau x} F_P(\tau ),
\end{equation}
where we used that for $x>0$
\begin{equation}
P\int_{-\infty}^\infty dy~ \frac{e^{iyx}}{y}=i\pi,
\end{equation}
and where we took $y=\tau-\omega$. Using Eq. (\ref{17}) we obtain
Eq. (\ref{fourierf}).

From Eq. (\ref{fourierf}) we see that the original
integro-differential equation (\ref{1}) is solved if we can solve
the Riccati equation (\ref{ricatti}), or, alternatively, the linear
second order differential equation (\ref{linear}).

\section{The case of no merging}

Before we discuss the Riccati equation it is quite instructive to consider
the case of no merging, $\beta =0$. In our previous work we have solved this
problem in terms of the Airy function
\cite{ferkinghoff,mathies} with the result
\begin{equation}
N(x)_{\beta =0}={\rm const.}~\int_{-\infty}^\infty d\omega~ \omega^2~
e^{i\omega^3/3+i\omega x}.
\label{free}
\end{equation}
This function solves the equation
\begin{equation}
{\cal O}(\omega)\tilde{N}(\omega)=0
\end{equation}
subject to the boundary condition that $N(x)\rightarrow 0$ for $x\rightarrow
\infty$,

With the present method we obtain from Eq. (\ref{ricatti}) with
$\beta =0$
\begin{equation}
-\frac{dF(\tau )}{d\tau}+\left(\frac{2}{\tau}+i\tau^2\right)F(\tau )-N'(0)-
2i\frac{c}{\tau}=0,
\label{linearapprox}
\end{equation}
where the constant $c$ is given by $c=\tilde{\beta}/\beta=\int_0^\infty dxN(x)
$. The solution is given by
\begin{equation}
F(\tau )=\tau^2 ~e^{i\tau^3/3}\left[K-\int \frac{d\tau}{\tau^2} ~e^{-i\tau^3/3}
\left(N'(0)+
2i\frac{c}{\tau}\right)\right].
\label{solution}
\end{equation}
Here $K$ is a constant of integration. Comparing to the correct result
(\ref{free}) by
use of the Fourier transform (\ref{fourierf}) in terms of $F$, we see that the
second term in the solution (\ref{solution}) apparently gives a deviation
from (\ref{free}).

The resolution of this paradox is that the last term in Eq.
(\ref{solution}) does not give any contributions to the Fourier
integral. This can be seen by noticing that the indefinite integral
in Eq. (\ref{solution}) can be expressed in terms of the incomplete
$\Gamma$ function, and when multiplied by the prefactor $\tau^2
~e^{i\tau^3/3}$ it is analytic in the upper half plane. When we plug
the second term in (\ref{solution}) into the Fourier integral
(\ref{fourierf}) we can close the integration by a large circle in
the upper half plane. This circle contributes nothing, since the
factor exp$(i\tau x)$ gives an exponential damping on the circle,
and since the behavior of the second term (including the prefactor)
in (\ref{solution}) is $-iN'(0)/\tau^2+O(1/\tau^3)$ for large
$\tau$, as can be seen by a partial integration. This behavior does
not compete with the  exponential damping. Therefore
\begin{equation}
-\tau^2 ~e^{i\tau^3/3}\int \frac{d\tau}{\tau^2} ~e^{-i\tau^3/3}
\left(N'(0)+2i\frac{c}{\tau}\right)
\label{null}
\end{equation}
is a null function in the Fourier integral (\ref{fourierf}).

The moral of this story is that although the function $N(x)$ is
given by the two integrals (\ref{fourier}) and (\ref{fourierf}),
this does not mean that the two functions $\tilde{N}$ and $F$ are
proportional. They can differ by a non-trivial null function. As the
example given above shows, we must in general expect the occurrence
of such non-trivial null functions.

Another point which should be mentioned is why the first term in the
solution (\ref{solution}) cannot be treated like the second term: In
the second term the factors exp$(\pm i\tau^3/3)$ cancel out, as one
can see by repeated partial integrations of the integral in the
solution (\ref{solution}). For example, for small $\tau$ one has
\begin{equation}
\tau^2~e^{i\tau^3/3}\int\frac{d\tau}{\tau^3}e^{-i\tau^3/3}=-\frac{1}{2}-
\frac{i}{2}\tau^2~e^{i\tau^3/3}\int d\tau ~e^{-i\tau^3/3}=-\frac{1}{2}-\frac{i}
{2}\tau^3+...~ .
\end{equation}
A similar argument does not apply to the first factor in
(\ref{solution}) which contains exp$(i\tau^3/3)$ and  does not give
convergence on the large circle: This factor overwhelms the damping
from exp$(i\tau x)$. Actually the contour can only be be closed in
the upper half plane by two straight lines in the directions $\pi
/6$ and $5\pi /6$. The closed contour running along the real axis
from $-\infty$ to $+\infty$, a part of a circle at infinity going
from the angle 0 to $\pi/6$ (gives no contribution due to a damping
factor exp$(-|\omega|^3 \sin (3\phi)),~0< \phi <\pi/6)$), runs back
to the origin along a straight line, and moves to the left along a
line making the angle $5\pi /6$, and is closed by a circle going
towards the real axis (gives no contribution). On the straight lines
the factor exp$(i\tau^3 /3)$ becomes strongly damped and behaves
like exp$(-|\tau|^3/3)$. Therefore the oscillating integral
(\ref{free}) can be replaced by the strongly damped integral
\begin{equation}
N(x)_{\beta=0}={\rm const}\int_0^\infty d\tau~\tau^2~e^{-\tau^3/3-\tau x\sin
\pi/6}~\sin (\tau x\cos \pi/6).
\label{watson}
\end{equation}

\section{The case of no diffusion}

We shall now consider the case without diffusion, where the Riccati
equation simplified to Eq. (\ref{nodiffusion}) with the
corresponding second order equation
\begin{equation}
 u''(\tau )-\left(\frac{2}{\tau}+2ib\right)u'(\tau )+\frac{2ib}{\tau}u(\tau )
=0.
\end{equation}
This equation can be solved in terms of Bessel functions,
\begin{equation}
u(\tau )=k\tau^{\frac{3}{2}}~e^{ib\tau}\left(J_{-\frac{3}{2}}(-b\tau )+
CN_{-\frac{3}{2}}(-b\tau )\right),
\end{equation}
where $k$ and $C$ are constants. These Bessel functions can
be expressed in terms of trigonometric functions,
\begin{equation}
 J_{-\frac{3}{2}}(-b\tau )=\pm i\sqrt{\frac{2}{\pi\tau b}}~\left(\sin (b\tau )
+\frac{\cos (\tau b)}{b\tau}\right),~~N_{-\frac{3}{2}}(-b\tau )=\pm i
\sqrt{\frac{2}{\pi\tau b}}~\left(\cos (b\tau )-\frac{\sin (\tau b)}
{b\tau}\right)
\end{equation}
For the function $F$ we then get
\begin{equation}
F(\tau)=ib~\frac{2f}{\beta}~\frac{(1+b\tau (C-i))\cos (b\tau)-(C-b\tau (1+iC))
\sin (b\tau)}
{(1+Cb\tau)\cos (b\tau)+(b\tau-C)\sin (b\tau)},
\label{sol}
\end{equation}
where $C$ is a constant of integration. Since $F$ is given by a Hilbert
transform, we see that for $\tau\rightarrow\infty$
\begin{equation}
F(\tau)\rightarrow \frac{1}{\tau}\int d\omega \tilde{N}(\omega)=\frac{1}{\tau}~
N(0),
\end{equation}
with correction terms starting with $-iN'(0)/\tau^2$. Thus, if $N(0)=0$, we
have that $F$ behaves like $1/\tau^2$. Going back to the solution
(\ref{sol}), we see that it is only possible to let $F$ go like $1/\tau$,
which fixes the integration constant to be $C=i$. Then
\begin{equation}
F(\tau)=\frac{2f}{\beta}~\frac{ib}{1+ib\tau}.
\label{pole}
\end{equation}
We can now perform the Fourier transform (\ref{fourierf}) by means of
Cauchy's theorem, and we obtain
\begin{equation}
N(x)=\frac{2f}{\beta}~e^{-x/b},
\end{equation}
It is easily verified that this function satisfies the original
integro-differential equation (\ref{1}) without the diffusion term.
The solution is identical to the one found in \cite{stewart}, by
noticing that in the case of no diffusion, $b$ can be expressed in
terms of the total ``mass'' $M=\int_0^\infty dx xN(x)$ from Eq.
(\ref{2b}) as $b=\sqrt{\beta/(2f) M}$. Clearly, this solution does not
satisfy $N(0)=0$. Instead there is a piling up at $x=0$ given by the
ratio of fragmentation versus merging. The more fragmentation one
has, the more piling up comes about for small $x$. Also, the more
merging one has, the less piling up at small $x$. These results are
of course physically reasonable.

\section{The effect of diffusion}

If diffusion is included, it will have the profound effect that it
is possible to enforce the boundary condition $N(0)=0$. We then have
to consider the full Riccati equation (\ref{ricatti}). Unfortunately
we have not been able to solve this equation exactly, in contrast to
the no diffusion case considered in the last section. Therefore we
need to find an approximate solution valid in different
$\tau-$ranges. In this connection it should be pointed out that if
$F(\tau)$ is (i) analytic in the upper half plane, and (ii) grows
less than exponential on a large semi-circle in the upper half
plane, then there is only the trivial solution $N(x)=0$. This simply
follows by the use of Eq. (\ref{fourierf}) by closing the contour in
the upper half plane at no price, since there is an exponential
damping from the factor exp$(ix\tau)$ due to (ii). Application of
Cauchy's theorem then gives the trivial result, since no
singularities are included inside this contour, due to (i).
Therefore, to have a non-trivial approximate solution at least one
of the two conditions (i) and (ii) must be violated.

To see that the boundary condition $N(0)=0$ is possible when diffusion is
included, we notice that for large $\tau$ the linear second order equation
(\ref{linear}) has the solution
\begin{equation}
u(\tau)\approx {\rm const.} \left(1+\frac{i\beta}{2fx_0\tau}~N'(0)+O(1/\tau^2)
\right),~~~{\rm i.e.,}~~\log u(\tau)\approx {\rm const.}+\frac{i\beta}{2fx_0\tau}~
N'(0)+O(1/\tau^2) ,
\end{equation}
which comes as a cancellation of the diffusive terms
$-i\tau^2u'(\tau)$ and $+\frac{\beta}{2fx_0} ~N'(0)$. Using Eq.
(\ref{log}) we have
\begin{equation}
N(0)=\frac{1}{2\pi i}\int_{-\infty}^\infty d\tau F(\tau)=\frac{2fx_0}{\beta}~
\frac{1}{2\pi i}\int_{-\infty}^\infty d\tau \frac{d}{d\tau}\log u(\tau )=
\frac{2fx_0}{\beta}~\left[\log u(\tau)\right]_{-\infty}^\infty =0.
\end{equation}
It should be noticed that in the case of no diffusion the asymptotic
behavior of $u(\tau)$ is different, $u(\tau)\approx {\rm const.}+O(\log \tau)$.

We shall now show that if one has an approximate pole behavior of $F(\tau )$
this represents a saddle point in the
Fourier integral (\ref{fourierf}). This is important, since we cannot directly
use Cauchy's theorem to perform the integral if
the solution is only approximate and dominates only near a pole $\tau_0$.
From the Fourier integral we see that there is a stationary phase (or
saddle point) when
\begin{equation}
ix\tau-\log(\tau-\tau_0)
\end{equation}
is stationary, which happens for
\begin{equation}
ix-\frac{1}{\tau-\tau_0}=0,~~{\rm i.e.}~~\tau=\tau_0+\frac{1}{ix}.
\end{equation}
Also, the second derivative of the phase $-x^2/4$ is negative, so the saddle
point is stable. The value of $N(x)$ in this saddle point is
\begin{equation}
N(x)\approx {\rm const.}~\sum e^{i\tau_0~x},
\label{saddle}
\end{equation}
where we should sum over all relevant $\tau_0$'s subject to the condition
that $N(x)$ is finite and positive.

We shall now investigate the possibility that there exists a pole
for small values of $\tau$. To this end we shall use that from Eq.
(\ref{log}) a pole in $F(\tau)$ can show up as a zero in the
function $u(\tau )$, which satisfies the linear second order
equation (\ref{linear}). Let us tentatively assume that this pole
occurs for small values of $\tau $, so that we can expand
\begin{equation}
u(\tau )\approx 1+c_1\tau +c_2\tau^2+...,
\label{2001}
\end{equation}
where the first constant is taken to be 1, since the overall scale
of $u(\tau )$ is irrelevant for $F(\tau )$. From (\ref{linear}) we then obtain
\begin{equation}
u(\tau )\approx 1+i\frac{b}{x_0}\tau+\frac{\beta}{2x_0f}~N'(0)\tau^2+...~.
\end{equation}
There is indeed a pole $u(\tau_0)\approx 0$,
\begin{equation}
\tau_0=\frac{i}{\beta/(fx_0) N'(0)}~\left(\sqrt{(b/x_0)^2+2\beta/(fx_0) N'(0)}-b/x_0\right),
\label{2000}
\end{equation}
where we took the root which leads to a finite result for $N(x)$ by use of
Eq. (\ref{saddle}). Using Eq. (\ref{2b}) this can be written
\begin{equation}
\tau_0=\frac{i}{<x>-b/x_0}~\left(\sqrt{\frac{2<x>}{b/x_0}-1}-1\right).
\label{slope}
\end{equation}
It is clear that this result is only valid provided $|\tau_0|$ is small,
since otherwise more terms should be included in the expansion
(\ref{2001})

Now the question naturally appears as to whether there are other
saddle points relevant when $\tau $ is large? In general, the saddle points
are in the
complex $\tau-$plane, as we saw in the pole case discussed above, where
$\tau_0$ is imaginary. In this connection we remind the
reader that in the case of no merging, we obtained as a solution an
oscillating integral (\ref{free}), which could be turned into an
integral (\ref{watson}) which is damped at large $\tau$ by deforming the
contour. On the deformed contour the Fourier transform is thus small
for large $\tau$. Motivated by these considerations we ask if there
exists an approximate solution of the Riccati equation (\ref{ricatti}) where
$F$ is small. In this case we can ignore the quadratic term,
and the equation reduces to
\begin{equation}
-\frac{dF(\tau )}{d\tau}+\left(\frac{2}{\tau}+i\tau^2+
\frac{2ib}{x_0}\right)F(\tau )-N'(0)-2i\frac{\tilde\beta}{\beta\tau}\approx 0.
\label{ricattilinear}
\end{equation}
We see that except for the term $\frac{2ib}{x_0}F(\tau)$ the
equation is the same as the equation with no merging
(\ref{linearapprox}), and the solution is the same as Eq.
(\ref{solution}), except for a contribution exp$(2ib/x_0 \tau)$.
Again there is a null function with respect to the Fourier integral
(\ref{fourierf}), similar to the one that occurs in
(\ref{solution}). Ignoring the null function we have
\begin{equation}
F(\tau)\approx {\rm const.}\tau^2~e^{\frac{i}{3}\tau^3+\frac{2ib}{x_0}\tau}.
\end{equation}
Along the contour forming angles $\pi/6$ and $5\pi/6$ with the real axis this
function is small for large values of $\tau$. The expression for $N(x)$
is then
\begin{equation}
N(x)\approx {\rm const.}~\int_{-\infty}^\infty d\tau \tau^2 e^{\frac{i}{3}
\tau^3+\frac{2ib}{x_0}\tau +ix\tau }.
\end{equation}
For large $x$ this leads to
a saddle point in the Fourier transform (\ref{fourierf})
well known
from the theory of Airy functions. It occurs at $\tau =i\sqrt{x}$ and is
stable,
\begin{equation}
N(x)\approx {\rm const.}~x^{\frac{3}{4}}~e^{-\frac{2}{3}x^{\frac{3}{2}}-
\frac{2b}{x_0}x^{\frac{1}{2}}}.
\label{airy}
\end{equation}

To conclude this discussion, we see that there are in principle
two saddle points (at least, there may be others that we have overlooked).
The one given by the slope (\ref{slope}) always dominates, because it decays
exponentially in contrast to (\ref{airy}). However,
there may be a transitory region for large, but not too large $x$,
where the behavior is given by (\ref{airy}), but for very large $x$ the
behavior is always dominated by the slope (\ref{slope}).
If the slope (\ref{slope}) increases to around one, the expression
(\ref{slope}) is no longer expected to be valid, since more terms
would be needed in the expansion (\ref{2001}).

\section{Numerical results}

The most direct way of validating the theoretical predictions
presented in the preceding section is to integrate Eq.
(\ref{fullequation}) numerically. Inevitably, a numerical approach
implies that the domain of the function $N(x)$ will be restricted to
a finite set of natural numbers $X=\{0,1,\cdots, L\}$. Here, we must
ensure that $L$ is chosen sufficiently large to suppress finite size
effects and that $X$ is sufficiently ``dense'' to give a proper
representation of the continuous function $N$. The second
requirement is satisfied by rescaling the diffusion to fragmentation
ratio so that $x_0^{-1}=\left(\frac{D}{f}\right)^{-1/3}\ll 1$.
Consequently, $L$ must be chosen such that $L/x_0\gg 1$. As we have
seen, the other relevant length scale is set by the coagulation to
fragmentation ratio $b=\frac{\beta}{2f}\int_0^\infty N(x)$. Since
diffusion processes will always be dominating for small values of
$x$, this length scale only plays a part in setting the upper limit,
i.$\!$ e. $L/b\gg 1$. The most important concern in choosing $L$ comes
from the problem of ensuring the mass conservation of Eq.
(\ref{fullequation}) $\partial_t \int_0^\infty dx N(x,t)x=0$. The
proper discretized version of the fragmentation and coagulation
operators that entails mass conservation in the finite domain $X$
reads:
\begin{equation}
[\partial_t N]_{\mbox{frag}}~~~\rightarrow~~~[\partial_t \tilde
N]_{\mbox{frag}}=-f(\tilde x-1)\tilde N(\tilde x)+2f\sum_{\tilde
  x'=\tilde x+1}^L \tilde N(\tilde x'),
\end{equation}
and
\begin{equation}
[\partial_t N]_{\mbox{coag}}~~~\rightarrow~~~[\partial_t \tilde N]_{\mbox{coag}}=\beta/2 \sum_{\tilde x'=0}^{\tilde x-1}\tilde N(\tilde x')\tilde N(\tilde x-\tilde x')-\beta \tilde N(\tilde x)\sum_{\tilde x'=0}^{L-\tilde x}\tilde N(\tilde x'),
\end{equation}
where $\tilde N$ represents the discretized function.
The discrete Laplacian
\begin{equation}
\partial^2_x N(x)~~~\rightarrow~~~\Delta^2
\tilde N(\tilde x)=\left\{
\begin{array}{ll}
\tilde N(1)-2\tilde N(0) & \mbox{for $\tilde x=0$} \\
\tilde N(\tilde x+1)+\tilde N(\tilde x-1)-2\tilde N(\tilde x) & \mbox{for $0<\tilde x<L$}, \\
\tilde N(L-1)-2\tilde N(L) & \mbox{for $\tilde x=L$}
\end{array} \right.
\end{equation}
however will always lead to a net loss of the total mass, since:
\begin{equation}
\sum_{\tilde x=0}^{L} \tilde x\Delta^2 \tilde N(\tilde x)= -(L+1)\tilde N(L).
\end{equation}
Here, we take the simplest approach of setting $L$ so that the
relative loss of mass will be small in each time step,
i.$\!$ e. $L\tilde N(L)\ll \sum_{\tilde x=0}^L \tilde x\tilde N(\tilde x)$.
Since for all parameters of $b/x_0$ we expect the distributions to
display an exponential decay (at least), the mass loss will
automatically be small when $L/x_0\gg 1$ and $L/b\gg 1$. In the
following we present results using $x_0\approx 50$ and $L=1000$ for
values of $b/x_0\le 2$ and $L=10000$ for $b/x_0>2$. The discretized
equation;
\begin{equation}
\partial_t \tilde N(\tilde x)=\Delta^2 \tilde N(\tilde x)+[\partial_t \tilde N]_{\mbox{frag}}+[\partial_t \tilde N]_{\mbox{coag}}
\end{equation}
is solved using fourth order Runge-Kutta integration scheme with
variable time-step \cite{numrec}. Convergence is obtained by requiring
that the relative change $\Delta {\cal N}$ of the total counts, ${\cal N}=\sum_{x=0}^L
\tilde N(x)$ during a time interval of $\Delta t=1$ satisfy $\Delta
{\cal N}/{\cal N}<10^{-6}$. For all calculations the time-integrated
mass loss is less than $0.1 \%$ compared to the mass of the initial
distribution.

\begin{figure}
\epsfig{width=.8\textwidth,file=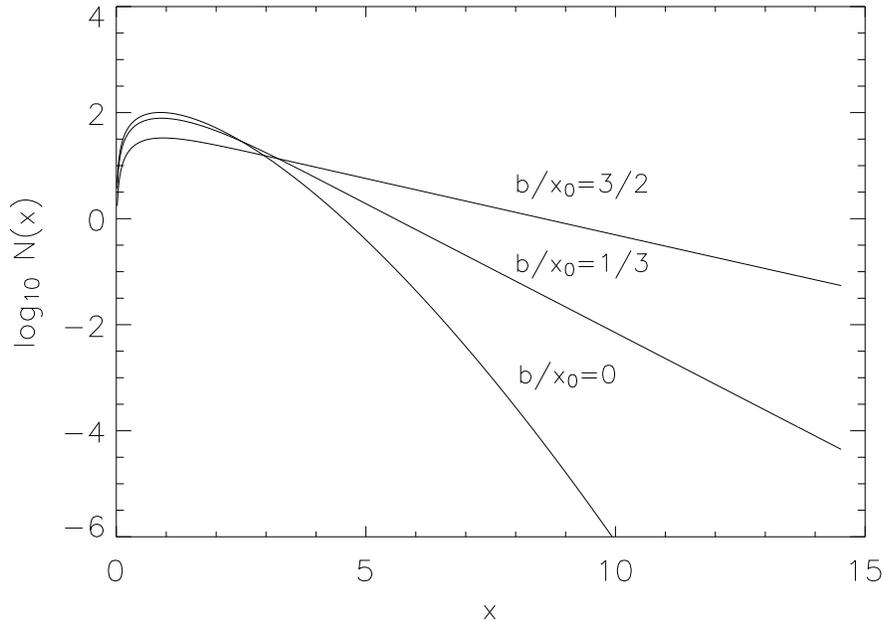} \caption{The numerical
steady-state distributions $N(x)$ as function of $x$ (both quantities 
dimensionless) for different
coagulation-to-fragmentation ratios $b/x_0$.}\label{distribution_fig}
\end{figure}

In Fig. \ref{distribution_fig} we show $N(x)=\tilde N(\tilde x)$ as
function of $x=\tilde x/x_0$ for different values of $b/x_0$. For
small values of $b/x_0$, one indeed observes a cross-over from the
functional form Eq. (\ref{airy}) at intermediate values of $x$ to a
pure exponential for large values of $x$. At larger values of
$b/x_0$, only a single exponential form is observed for $x>1$.

\begin{figure}
\epsfig{width=.8\textwidth,file=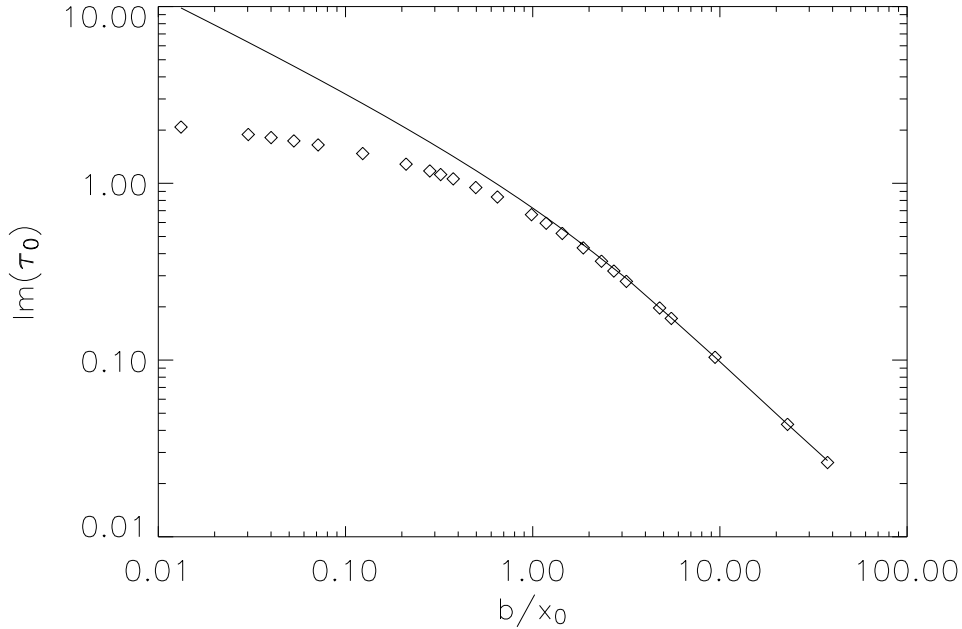} \caption{The slope
$Im(\tau_0)$ (dimensionless) of the exponential tail of $N(x)$ as function of
  $b/x_0$. The solid line is the theoretical prediction, Eq. (\ref{slope})
  and the square diamonds are the numerical results. One observes a
  perfect agreement for all $Im(\tau_0)<1$.}\label{slope_fig}
\end{figure}

In Fig. \ref{slope_fig} we show the slope $Im(\tau_0)$ of the
exponential tail of the distributions as function of $b/x_0$. The
full curve is expression (\ref{slope}), whereas the square diamonds
are the slopes obtained from an exponential fit to the tail of the
numerical solutions for different values of $b/x_0$. For
$\Im(\tau_0)<1$, which corresponds to values where the coagulation
to fragmentation ratio is the dominant length scale, $b>x_0$, we
find perfect agreement between the theoretical predictions and the
numerical results.


We shall now use a simple 1D lattice model as an alternative
approach to put the aforementioned theory to a test. We pick a
finite lattice of length $L$ and impose periodic boundary
conditions. On the lattice we initially distribute a set of
$\mathcal N$ clusters of sizes $\{\ell_i\}_{i=1}^\mathcal N$ and
with a total mass chosen to fit the lattice length,
$\sum_i\ell_i=L$. We define the cluster boundaries by a new set
$\{x_i\}_{i=1}^\mathcal N$ such that $\ell_i=x_i-x_{i-1}$ (with
$x_0$=$x_N-L$). The diffusion process of Eq. (\ref{fullequation})
corresponds in the lattice model to a random walk of the boundaries
and therefore in each time step we update the boundary positions
according to $x_i(t+1)= x_i(t)\pm1$. Here the lattice spacing and
time steps, as well as the diffusion constant, are chosen to be unity.
A fragmentation event is simulated by introducing a new random
walker or boundary $\tilde x$ on the lattice. The position of
$\tilde x$ is picked uniformly among the $L-\mathcal N$ available
lattice sites. Effectively, we therefore get a constant
fragmentation rate given by $f=n/L$, where $n$ is the number of new
boundaries introduced in each time step. Finally, the coagulation
process is similar to the removal of boundaries. In each time step a
random walker is removed with a probability $\gamma$. Note that
$\gamma$ is equivalent to the rate $\tilde \beta=\beta \mathcal N$
introduced in (\ref{tildebeta}). The equivalence follows by
comparing the number of events where a cluster of
size $x$ coagulates with a cluster of size $y$. In the lattice model,
the number of events in a unit time are $\mathcal N
\gamma\frac {N(x)}{\mathcal N} \frac {N(y)}{\mathcal N}$, which should
be compared with the corresponding number in Eq.
(\ref{fullequation}), and thus we get $\gamma/\mathcal N=\beta$.
\begin{figure}
  \epsfig{width=.5\textwidth,file=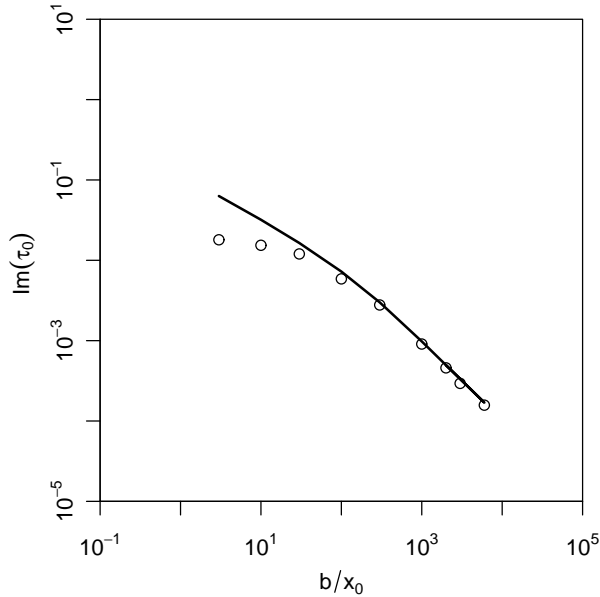}
  \caption{Lattice simulation data of $\mbox{Im}\,\tau_0$ as function of $b=\tilde\beta/f$. The
line represents the analytical expression of $\mbox{Im}\,\tau_0$.
}\label{DFCP1}
\end{figure}
In order to minimize the effect of the underlying lattice it is
important to keep the total mass or lattice length $L$ of the system
much larger than any other scale in the system, i.$\!$ e. $L\gg
(D/f)^{1/3}$ and $L\gg \tilde\beta/f$. Fig. \ref{DFCP1} presents
estimates of (\ref{slope}) based on the lattice model with a lattice
size $L=10^7$. We collect the simulation data in a histogram with a
unit bin-size and then fit the exponent of the tail. The maximum
count in any bin is for reasonable amounts of computer time limited
above by $10^8$, thus we have a bound on the maximum exponent that it is
possible to estimate and we have an explanation why the
lattice simulations provide poor estimates in Fig. \ref{DFCP1} for
low values of $b$.

Compared to the direct numerical integration there are both
advantages and disadvantages. First of all the lattice model is
extremely simple from a computationally point of view and does by
construction conserve the total mass. Moreover the model reproduces
to high accuracy, using little computer time, the theoretical
predictions for large values of $b/x_0$, which is in contrast to the
larger computational power needed in the direct numerical
integration. Note that in the lattice model there is a weak size
correlation between neighboring clusters. The diffusion makes some clusters
large on the cost of the surrounding ones. The correlation is not
present in the mean field equation (\ref{fullequation}) and it may
be avoided on the lattice by shuffling in each time step the
clusters. By implementing this shuffling in the numerical
routine, it however turns out, that the correlation
has negligible or no effect on the cluster size distribution.

\section{CONCLUSION}

In this paper, we have introduced a model which includes the three
fundamental physical processes: diffusion, fragmentation and
coagulation. The model is formulated as a dynamical equation in
terms of the distribution $N(x,t)$ of fragment sizes $x$. The main
results of the paper are: In the case of no coagulation term, we
obtain an exact solution for the distribution of the Bessel type
${\rm exp}(-\frac{2}{3}x^{3/2})$. When the coagulation terms is
added, we show that the non-linear equation can be mapped exactly
onto a Riccati equation. From solution of this equation we
obtain that the distribution now turns into a pure exponential for
large $x$. When the coagulation process is small as compared to the
fragmentation process, we identify directly that the distribution
$N(x)$ behaves as the Bessel function for small $x$ after which it
crosses over to an exponential at a specific value of $x$.

We believe that our proposed model is relevant in many physical
situations, such as for solutions of macromolecules like polymers,
proteins and micelles. In fact, from measured distributions of
fragment sizes we suggest that one might be able to identify how
important the coagulation process is compared to the fragmentation
process. We are in the process of collecting experimental data for
such investigations.

\end{document}